\documentclass[a4paper,11pt]{article}
\usepackage{pos}
\usepackage{url}

\usepackage{natbib}
\usepackage{adjustbox}
\usepackage{subfig}
\usepackage{booktabs}

\title{Gauge generation and dissemination in OpenLat}

\dedicated{
\phantom{.}\\
\vspace{6ex}
\includegraphics[width=0.3\textwidth]{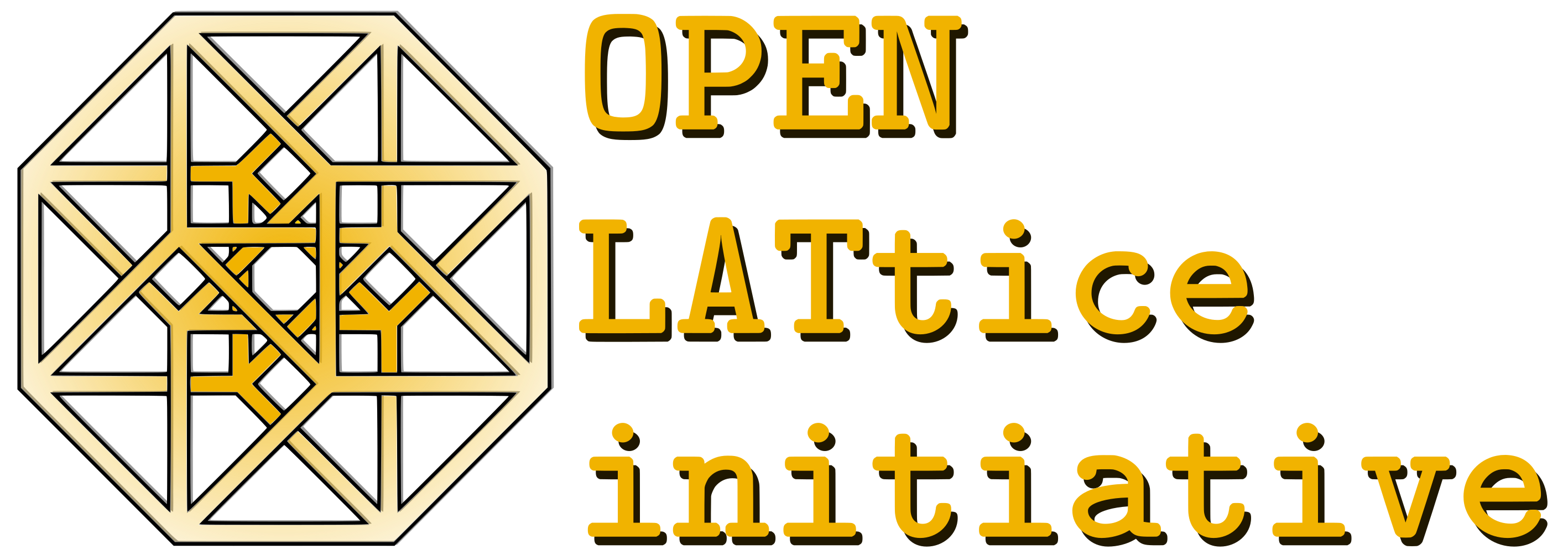}
}

\author[a]{F. Cuteri}
\author*[b]{A. Francis}
\author[c]{P. Fritzsch}
\author[d]{G. Pederiva}
\author[e,f]{A. Rago}
\author[g]{A. Shindler}
\author[h]{A. Walker-Loud}
\author[i]{S. Zafeiropoulos}

\affiliation[a]{Institut für Theoretische Physik, Goethe Universität, 
Max-von-Laue-Str. 1, 60438 Frankfurt, Germany}

\affiliation[b]{Institute of Physics, National Yang Ming Chiao Tung University, 30010 Hsinchu, Taiwan}

\affiliation[c]{School of Mathematics, Trinity College Dublin, Dublin 2, Ireland}

\affiliation[d]{J\"ulich Supercomputing Centre, Forschungszentrum J\"ulich GmbH, 52428 J\"ulich, Germany}

\affiliation[e]{IMADA \& $CP^3$-Origins, University of Southern Denmark, Odense, Denmark}
\affiliation[f]{Theoretical Physics Department, CERN, CH-1211 Geneva 23, Switzerland.}

\affiliation[g]{FRIB \& Physics Department,
Michigan State University, East Lansing, MI 48824, USA}

\affiliation[h]{Nuclear Science Division, Lawrence Berkeley National Laboratory, Berkeley, CA 94720, USA}

\affiliation[i]{Aix Marseille Universit\'e, Universit\'e de Toulon, CNRS, CPT, Marseille, France.}

\abstract{In this contribution we report the status and plans of the open lattice initiative to generate and share new gauge ensembles using the stabilised Wilson fermion framework. The production strategy is presented in terms of a three stage plan alongside summaries of the data management as well as access policies. Current progress in completing the first stage of generating ensembles at four lattice spacings at the flavor symmetric point is given.}

\FullConference{%
  The 39th International Symposium on Lattice Field Theory (Lattice2022),\\
  8-13 August, 2022 \\
  Bonn, Germany 
}

\begin{document}
\maketitle

\section{Introduction}

The current high precision era of lattice QCD introduces new questions on data generation, availability and access. With the increase in precision comes also an increase in numerical cost as well as the size and complexity of the data. Especially data sharing and community access become important as the process of transferring data can be a problem in itself and not having access can prohibit impactful physics analyses. 
Simultaneously, the current level of precision is built on a wealth of previous data driven studies. In principle progress can be slowed by having to invest first to re-establish a comparable foundation for a new method. 
In this situation it is important as community to re-evaluate the existing data sharing infrastructure, e.g. in form of the ILDG, and develop an expanded understanding of open science as a tool for scientific progress.

\section{The open lattice initiative, setup and features}

The open lattice initiative (OpenLat) was established as a response to some of these questions. Bringing together researchers across the globe, the goal is to generate state-of-the-art QCD gauge ensembles for physics applications and to share them with the community to strengthen open science.
As part of this effort the initiative aims to transparently define and uphold quality for the configurations provided (1), share and maintain an easily accessible repository (2), gather resources towards a common aim (3), and to grant access to any interested party freely and quickly (4).

The stabilised Wilson fermion (SWF) framework \cite{Francis:2019muy} includes a modified improvement term for the fermion action and also a number of advancements aimed at algorithmic stability\footnote{The SWF framework is also being used in the context of the master-field approach, see \cite{pfritzsch:lattice22} for an overview.}.
These measures are implemented in addition to the established techniques implemented within the open source software package \texttt{openQCD-2.0} and \texttt{2.4} \cite{mluscher:openqcd}, such as the Schwarz-alternating-procedure (SAP), local deflation, mass-preconditioning, multiple time-scale integrators and others. 

All generated gauge ensembles are in $N_f=2+1$ QCD. To set the scale the gradient flow time criterion \cite{Luscher:2010iy} is used and results are converted to physical units via $\sqrt{8t_0}=0.414(5)\textrm{fm}$ \cite{Bruno:2016plf} .

\subsection{Stabilised Wilson Fermions}
\subsubsection{Exponentiated Clover action}

An important new component in OpenLat is the inclusion of a new fermion action resulting from a modification of the standard Wilson clover (WCF) action. The (traditional) $\mathcal{O}(a)$-improved Dirac operator takes the form
\begin{equation}
D = \frac{1}{2} \Big[\, \gamma_\mu\Big( \nabla_\mu^*+\nabla_\mu \Big)\,- \nabla_\mu^*\nabla_\mu \Big] + c_{SW} \frac{i}{4}\sigma_{\mu\nu}\hat F_{\mu\nu} + m_0~~,
\end{equation}
which, after even/odd preconditioning, may be written as $\hat D = D_{ee} - D_{eo} ( D_{oo} )^{-1} D_{oe}$, with a diagonal part ($M_0=4 +m_0$) given by
$D_{ee} + D_{oo} = M_0 + c_{SW} \frac{i}{4} \sigma_{\mu\nu} \hat F_{\mu\nu} $. Notice that arbitrarily small eigenvalues can occur as the clover term can saturate the bound
$\Vert \frac{i}{4} \sigma_{\mu\nu} \hat F_{\mu\nu} \Vert_2 \leq 3$
while the clover coefficient $c_{SW}$ is one at tree-level and then grows monotonically with $g_0^2$.
Furthermore, the positive and negative eigenvalues of the clover term are equally distributed.
Taken together this makes the above statement more precise as we see that $D_{oo}$ is not protected from arbitrarily small eigenvalues. The probability to encounter arbitrarily small eigenvalues of $D_{oo}$ is larger, the smaller the quark masses, and/or the coarser/larger the lattice.
This issue can be resolved by reformulating the clover term:
\begin{equation} 
D_{ee} + D_{oo} = M_0 + c_{SW} \frac{i}{4} \sigma_{\mu\nu} \hat F_{\mu\nu} ~\rightarrow~~ M_0\exp\Big[ \frac{c_{SW}}{M_0}\frac{i}{4} \sigma_{\mu\nu} \hat F_{\mu\nu}  \Big]~~,
\end{equation}
this is the so-called exponentiated clover \cite{Francis:2019muy}. 
This formulation is still invertible, local and has cutoff effects described, as usual, by a Symanzik effective theory. While the lattice eo-preconditioned Dirac operator
is still not protected from exceptionally small eigenvalues stemming from the Wilson term, this new formulation eliminates the clover term as a potential
source of instabilities.

\subsubsection{Algorithmic features}

Part of the SWF framework is to include a number of algorithmic improvements, one of them is choosing the Stochastic Molecular Dynamics (SMD) algorithm \cite{Horowitz:1985kd,Horowitz:1986dt,HOROWITZ1991247,Jansen:1995gz} over the commonly used HMC. Benefits and features of the SMD were presented in \cite{Luscher:2011kk,Luscher:2017cjh}. Here we note that the algorithm leads to an increased stability and decreased autocorrelation times by reducing unbounded energy violations $|\delta H|\gg 1$.
The HMC and SMD are closely related, for example the HMC is a limiting case of the SMD for large step lengths and friction coefficients. In the opposite direction the SMD becomes related to a Langevin process.
As such, in an SMD update cycle the momentum fields are randomly rotated as opposed to randomly re-drawn and an update length a fraction of the typical trajectory length is chosen. 
One special feature of the SMD is that once there is rejection the momentum tends to reverse and trajectories backtrack with a period $t_{acc}=\delta \tau P_{acc}/(1-P_{acc})$ \cite{Luscher:2011kk}. This makes a very a high acceptance rate necessary.
Further improvement in stability can be achieved by changing the solver stopping criterion a volume-independent norm $\Vert \eta \Vert_\infty=\textrm{sup}_x \Vert \eta \Vert_2$ that prevents precision losses from local effects.
Finally, to limit the effect of an accumulation of numerical errors on global sums quadruple precision is implemented and used.

\subsection{Tuning Setup}


The clover coefficient $c_{SW}$ is tuned according to the procedure followed in \cite{Luscher:1996ug}. It is performed in the range $\beta\geq 3.685$ \cite{Francis:2019muy,Francis:2022hyr} corresponding to a coarse lattice spacing of $a=0.12$~fm. The resulting interpolation formula is given by
\begin{equation}
c_{SW}(g_0^2) = \frac{1-0.325022g_0^2-0.0167274g_0^4}{1-0.489157g_0^2}, \qquad \beta=6/g_0^2\geq 3.685.
\end{equation}


Special attention needs to be given to choosing the bare mass parameters for fixing the trajectory to physical masses and the continuum limit with Wilson fermions. We adopt a strategy that aims at reducing mass-dependent cutoff effects by working at constant trace of the simulated quark mass matrix \cite{Bietenholz:2010jr, Bruno:2014jqa,Strassberger:2021tsu}.
Starting at the flavour-symmetric point, where all pseudoscalar meson masses are degenerate, we set
$
    \textrm{tr}[M] = m_u + m_d + m_s = N_f \cdot m_\ell = \textrm{const} ~,
$
such that it depends on a single mass parameter. The value of $m_\ell$ amounts to matching the combination of the physical ground state masses of the pion and the kaon:
$    m_{\pi K}^2 = \frac{2}{3}\Big( m_K^2 + m_\pi^2/2\Big) \equiv m_{\pi K}^2|_{\textrm{phys}}$.
Following the procedure outlined in \cite{Aoki:2016frl} the input value is $m_{\pi K}|_{\textrm{phys}}=410.9(2)\,\rm{MeV}$.
Once the starting point is tuned and fixed, the quark masses are changed such that $\textrm{tr}[M]=\textrm{const}$.

\subsection{Quality control strategy}\label{sec:validation}

\subsubsection{Run stability and health}
During the generation of the gauge configurations a number of standard observables is continuously monitored and sanity checks are performed to test the health of the run. A few prominent validation observables are: the plaquette, topological charge, $\Delta H$ distribution, reweighting factors, distributions of the lowest eigenvalue $\lambda$ of $\sqrt{D^\dag D}$, and simple spectroscopic observables.
In particular, our checks are meant to ensure that\footnote{The last two items in the following complement the list we made in earlier references \cite{Francis:2022hyr}. In the future we plan to add the sign of the reweighting factor to the standard observables}:
\begin{itemize}
\itemsep0em
    \item $\phi_4=8t_0( m_K^2 + m_\pi^2/2)=1.115$ within $0.5\%$, with an error of max. $1\sigma$. 
    \item The total reweighting factor fluctuations are mild, and ideally below $5\%$. 
    \item The SMD step distance $\delta \tau$ maximises the backtracking period. 
    \item The distribution of $\delta H$ matches the one set by the acceptance rate. 
    \item The distribution of the lowest $\sqrt{D^\dag D}$ eigenvalue is well-behaved and gapped.
    \item\label{item:poles1} The distribution of the lower/upper bounds of the spectral gap for the strange quark are within the input ranges, and the degree of the Zolotarev is sufficiently high, $12 (V/2) \delta^2 < 10^{-4}$ \cite{mluscher:openqcd}.
    \item\label{item:poles2} There is no significant loss of precision caused by unbalanced contributions to the total action that might drive instabilities in the evolution.
    \item\label{item:topCharge} The distribution of the flowed topological charge is symmetric around zero with no signs of metastability.
\end{itemize}

\subsubsection{Distance between configurations and thermalisation}

Estimating the distance between two configurations that can be safely labelled as independent is based on evaluating the integrated autocorrelation times of several observables and choosing the slowest observed one. In most cases this is the topological charge $Q$, which is computed using the gradient flow \cite{Luscher:2010iy}. However, if another scale is seen to be even slower the distance is switched to this value instead. 
The boundary conditions are periodic/anti-periodic by default. However, once a significant increase of the autocorrelation time of $Q$ is observed the boundaries are changed to open boundary conditions. The increase is interpreted as a signal of entering a regime of topology freezing and critical slowing down. 
Currently this is only the case for $\beta=4.1$ and $a=0.055$~fm. All ensembles reported on in this proceedings contribution do not exhibit this feature and have periodic/anti-periodic boundary conditions.

The plaquette, $t^2\langle E\rangle$, where $t$ is the flow time and the density $E$ is defined in \cite{Luscher:2010iy}, pseudoscalar-pseudoscalar ($PP$) and pseudoscalar-axial ($PA$) meson correlators are checked for visible effects in order to estimate sufficient thermalisation. Nevertheless, a minimum of five integrated autocorrelation lengths, of the slowest scale, is imposed. This naturally relies on an accurate determination of the autocorrelation lengths which become available only after sufficiently long MC strings have been generated.

\subsubsection{Tuning and production levels}
\label{sec:tuningAndProd}

The gauge generation process is separated into two levels: The first is where thermalisation performed and the parameters of the algorithm tuned or adjusted. Once the algorithm is stable and thermalised, 100 independent configurations are generated. This is called the tuning level and once the health of the run is confirmed it is deemed production ready. At this point the initial 100 configurations become part of the full ensemble statistics.
At the production level configurations are produced with all run parameters fixed. Once the production reaches 500+ independent configurations in an ensemble it is labelled publication ready.

\section{Gauge generation and dissemination status}

\subsection{Production, publication and access plan}
\label{sec:plan}

The production of ensembles will proceed in three stages:
\begin{itemize}
\itemsep0ex
\item [{\bf 1.}] {\bf Stage:} Perform high precision tuning and generate ensembles with 3 dynamical flavors at the SU(3) flavor symmetric point. The minimum goal is to generate 500+ independent configurations on 4+ lattice spacings ($a=0.12, 0.094, 0.077$ and $0.064$~fm). 
\item [{\bf 2.}] {\bf Stage:} Reduce the light quark masses with Tr$[M]=$const. The minimum goal is to have matched pion passes at $M_\pi=300, 200$~MeV for $a=0.12, 0.094, 0.077$ and $0.064$~fm. An additional lattice with $a=0.055$fm (open boundary conditions) will be added at this stage.
\item [{\bf 3.}] {\bf Stage:} Go towards the physical values of the pion mass $m_\pi=135$~MeV. The minimum goal is to supply at least one ensemble at the physical point with 500+ configurations and then to extend towards all lattice spacings.
%
\end{itemize}
Multiple lattice volumes are generated at all stages and are extended towards 500+ configurations, where reasonable (i.e. $L\gtrsim 3$fm, $m_\pi L\gtrsim 4$).

Each completed stage is accompanied by a reference publication. With this publication all configurations and metadata of the given stage are made openly available without further embargo time. The data that will be made public are a metadata catalogue, the gauge configurations themselves and additionally the standard, or validation, observables highlighted above.
Details on how and where to download the configurations and their corresponding auxiliary data will be given at the same time. 
The metadata catalogue will be made compliant with further community standards, such as ILDG, to facilitate data sharing and a common data standard within the community.

Users may obtain access to the configurations of ongoing, i.e. unpublished, stages. This user-access is granted on a case-by-case basis\footnote{Users have reported results using OpenLat ensembles already at this and other conferences in \cite{fjoswig:lattice22,jgreen:lattice22,jkim:ectstar22}.}. 

\subsection{Data management plan}

Currently OpenLat maintains a repository of 100TB of data. This includes $\sim20$k saved configurations and is projected to grow to 500+TB with completion of stage 2. The repository is mirrored at currently two separate locations to ensure redundancy and a tape option is used for long term storage. All metadata is preserved on disk and in the main online repository at \href{https://openlat1.gitlab.io}{https://openlat1.gitlab.io}. Currently this access is only internal, it will however be made open upon completion of stage 1. The internal metadata follows a detailed provenance policy that includes run and machine information along side the previously mentioned standard observables.

To maintain data integrity a two part strategy is followed: the configurations contain the plaquette in the header, as per openQCD format standard, and is augmented by a separately kept list of checksums of sufficient complexity (e.g. \texttt{sha512}).

\begin{table}[t!]
    \centering
    \begin{tabular}{ccccrrccccr}
         \toprule
         Stage & $m_\pi$[MeV] & $\beta$ & $a\rm{[fm]}$ & {$L$} & {$T$}& BC & {$N_{\rm{cfg}}$} & MDU \\\hline
         1 & 412 & 3.685 & 0.12 & 24 & 96 & P & 1200 &  24800\\
         & &  &  & 32 & 96 & P & 400 &  8300 \\
         & & 3.8 & 0.094 & 24 & 96 & P & 1200 &  35000\\
         & & & & 32 & 96 & P & 1300 &  38800\\
         & & 3.9 & 0.077 & 48 & 96 & P & 300 &  11000 \\
         & & 4.0 & 0.064 & 48 & 96 & P & 600 &  24500\\ \hline
         2 & 300 & 3.685 & 0.12 & 24 & 96 & P & 600 & 13000\\
         & & 3.8 & 0.094 & 32 & 96 & P & 300 &  8000 \\
         & & 4.0 & 0.064 & 48 & 96 & P & 200 &  6000 \\ \hline
         \toprule
    \end{tabular}
    \caption{Stage 1 and 2 production status overview: All ensembles are currently still gathering more statistics and numbers should be regarded as preliminary. See Sec.\ref{sec:plan} for details on planned ensembles and target statistics. Note that the numbers in MDU have been derived by converting from SMD trajectory length.
}
\label{tab:ensembles}
\end{table}

\section{Production status}

The initiative is currently in the process of completing stage 1 of the production plan while stage 2 is being pursued, see Sec.~\ref{sec:plan}. An overview of the current number of configurations and ensembles available at the production level is given in Tab.~\ref{tab:ensembles}. 
The information on the MC chain length in MDU was obtained by multiplying the SMD step length with the total number of cycles.

\begin{figure}
\centering
\includegraphics[width=0.69\textwidth]{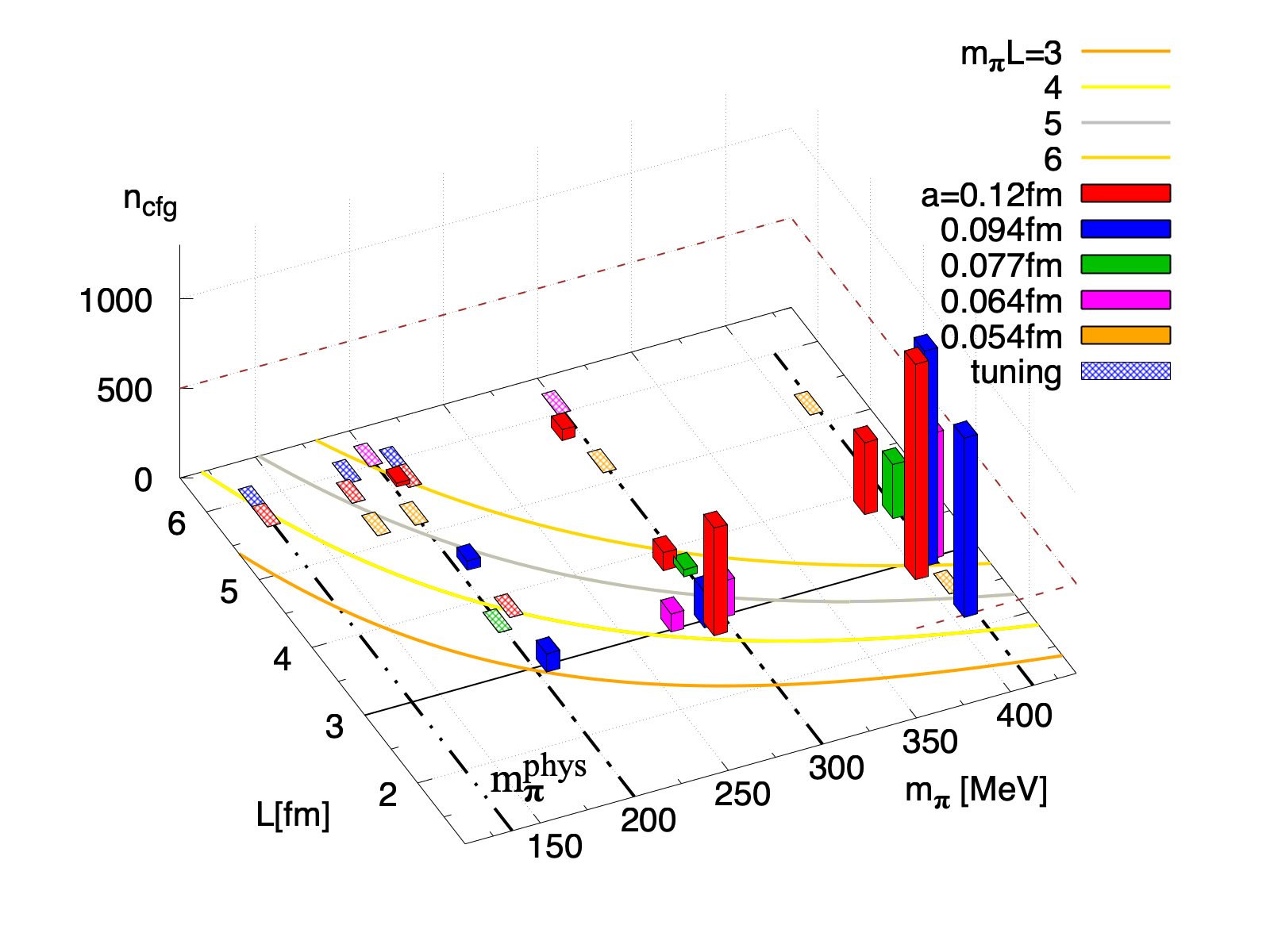}
\caption{Overview of ensembles at the tuning (shaded boxes) and production (solid boxes) levels. The shaded or colored lines are to highlight: Lines of constant $m_\pi L$, values of $m_\pi$[MeV], lattice volumes $L$ and the target number of configurations.}
\label{fig:overview}
\end{figure}

An overview of the full production landscape is given in Fig.~\ref{fig:overview}.
The figure shows the sets of pion masses and lattice spacings that are at the tuning level, i.e. have less than 100 independent configurations or are being otherwise thermalised or tuned, in shaded boxes. The solid boxes show those ensembles that are at the production level. Their height denotes the number of configurations they have.
The colored curved lines denote lines of constant $m_\pi L$, while the grey straight line denotes $L=3$~fm. The dashed floor lines are to guide the eyes as to the location of the target values in $m_\pi$[MeV]. The width and depth of the boxes have been chosen for legibility and do not reflect errors in $m_\pi$ or $L$. 
Note that 500 configurations for all $SU(3)$ flavor points is threshold for the completion of stage 1 and also making the ensembles openly accessible.

The focus on performing a high precision tuning and providing high statistics ensembles at the flavor symmetric point is clearly visible. Furthermore the newest lattice spacing, $a=0.077$~fm, stands out for being the lowest in statistics. With respect to a previous version of this figure we have removed an older physical point ensemble at $a=0.094$~fm for having a too small volume with $m_\pi L\simeq 2.1$.

\section{Summary}

In this contribution, we highlighted the recent progress made by the open lattice initiative. The initiative was founded with the expressed purpose of generating and providing ensembles for physics applications under open science. The ensembles are generated using the stabilised Wilson fermion framework, which is further researched and refined as part of the effort. 

The three stage production plan has been laid out and an overview of the production status given. As stage 1 is nearing completion, once its reference publication is published the plan is that access will be granted to the configurations under the presented data management and access plans. 

The procedures and results presented in this contribution should be regarded as preliminary and will be superseded by their final versions in the upcoming publication.

\section*{Acknowledgements}
OpenLat acknowledges support from the HPC computing centres hpc-qcd (CERN), HPE Apollo Hawk (HLRS) under grant stabwf/44185, Cori (NERSC), Frontera (TACC), Piz Daint (CSCS), Occigen (CINES), Jean-Zay (IDRIS) and Ir\`ene-Joliot-Curie (TGCC) under projects (2020,2021,2022)-A0080511504, (2020,2021,2022)-A0080502271 by GENCI and PRACE project 2021250098. This work also used the DiRAC Extreme Scaling service at the University of Edinburgh, operated by the Edinburgh Parallel Computing Centre on behalf of the STFC DiRAC HPC Facility. DiRAC is part of the UK National e-Infrastructure.
AS acknowledges funding support under the National Science Foundation grant PHY-2209185. AF acknowledges support under the Ministry of Science and Technology Taiwan grant 111-2112-M-A49-018-MY2. We greatly acknowledge the leading role of Martin L\"uscher during the development and implementation of the SWF framework.

\bibliographystyle{JHEP}
\bibliography{references}


\end{document}